# Overview of digital health surveillance system during COVID-19 pandemic: public health issues and misapprehensions

Molla Rashied Hussein[1] [*], Ehsanul Hoque Apu[2,3], Shahriar Shahabuddin[4],
Abdullah Bin Shams[5], and Russell Kabir[6]

[1] Department of Computer Science and Engineering, University of Asia Pacific, Dhaka, Bangladesh.
[2] Institute of Quantitative Health Science (IQ), Department of Biomedical Engineering, Michigan State University, East Lansing, MI 48824, USA.
[3] Laboratory of Developmental Biology, Faculty of Biochemistry and Molecular Medicine, University of Oulu, Aapistie 5A, Oulu 90220, Finland.
[4] Mobile Networks, Nokia, Oulu, Finland
[5] Edward S. Rogers Sr. Department of Electrical and Computer Engineering, University of Toronto Toronto, Canada.
[6] School of Allied Health, Faculty of Health, Education, Medicine and Social Care, Anglia Ruskin University, Chelmsford, Essex, UK.

*Corresponding author: Molla Rashied Hussein; Email: mrh.cse@uap-bd.edu

Introduction:

The novel coronavirus disease 2019 (COVID-19) spread globally and initially hit even the European countries with developed healthcare systems [1]. Without COVID-19 specific vaccine or medicine to prevent or cure, currently, the best approach is not to get exposed to the virus. Having close contact with someone, approximately less than six feet increases possibility of inhaling transmissible droplets from an infected person, through coughing, sneezing or talking. Social distancing must be maintained between everyone, as studies revealed that asymptomatic people may spread the disease [2].

The COVID-19 shows the necessity of automated contact tracing to track down newly infected cases, control, and predict the real time spreading. Contact tracing identifies unreported and asymptomatic infected individuals by tracing back who could have possibly exposed to an infected individual. Some countries developed systems without considering user privacy, whereas many followed the Singapore model based on Bluetooth data exchanges on contact [3].

Literature Review:

Numerous digital health surveillance systems (DHSSs) are used in several countries to identify infected people, observe mobility, and narrow down the tranmission risks. DHSSs also aid to prevent mass-level infections and an exponential number of untimely deaths, and build proactive recovery approaches with actions. Artificial intelligence (AI), big data, 5G technology team up with other next-generation devices, such as drones, automated vehicles, and robotics [4]. However, authorities need to be ethical and cautious regarding privacy and security, while appling DHSS for preventing public health mishaps.

China has extensively used DHSS, initially in transport system to track down the vehile movement and related information. Drones and 5G were used in monitoring the transport system during emergency period. Thermal cameras synced in the police helmets and public officials' head caps helped quick thermal screening in Guangdong, and the data was transmitted using 5G. Robotics with 5G helped sanitizing Wuhan in its peak period when manual services became risky. Similarly, a combination of automated vehicles and 5G was used for delivering goods in highly contaminated areas. For providing telemedicine care and pieces of advice at the brand-new hospital constructed in Wuhan, 5G was also utilized [4].

Hua and Shaw reported a health barcode system to trace the affected people in China [5]. Primarily, Hangzhou city used the system in early February, followed by two hundred other Chinese cities afterwards. For registration, a user must sign up for the app and scans a Quick Response (QR) code on the smartphone [4]. The app alerts the users of having proximity to an infected persons. The barcode system comprises three colors for coding: green for good health, yellow for precaution, and red for the infected person. These colors enable or disable users from entering different public buildings and mass transports.

Although Chen argued that the likelihood of receiving both positive and negative outcomes by using DHSS is quite high. It was discussed how a person responds under pressure; therefore, DHSS tools and epidemic maps require to be aligned accordingly [6]. Also, as big data are decisive, the data collection should be accountable, and comprehensible. If the algorithms are not used in an answerable approach, if the rules and regulations of data protection have not complied, and if the privacy, security, and confidentiality are not valued, then public trust will be shattered. Consequently, people will not follow seriously any guidance or suggestion provided by health authorities and will become more vulnerable to public health misapprehensions, and achieve poor outcome [7]. That means DHSS mishandling will shatter health system, and altered outcome may emerge from the initial projected one.

Internet of Things (IoT) creates a common platform to access data by the public health agencies to observe COVID-19 pandemic [8]. Viral activity modelling studies are possible to be implemented using big data. Public health policymakers can be directed to improve the outbreak preparation by big data. Social media can augment education and communication regarding public health, Singapore government has signed an agreement with WhatsApp to transmit precise COVID-19 information, and government initiatives to citizens. Numerous social media platforms, e.g., Facebook and Twitter, assist healthcare agencies in providing real-time updated information. Facial recognition companies, SenseTime and Sunell, have improvised the thermal imaging to facilitate recognition based detecting people with elevated temperature at several

Chinese screening points. Authorities in Italy, China and India have implemented experiments with fever-finding drones for a large scale screening and identifying infected persons. Draganfly and Quint have retrofitted drones with thermal cameras to measure high temperature from 100 meters distance, where the significant challenge is data accuracy. The FDA recommends that the fever sensing thermal devices should have an uncertainty of +/- 0.5°C. A study showed that drone based thermographic camera has an accuray of +/- 5°C [9], which attributed to wind and temperature change. Fourth, AI and deep learning can play a vital role in improving COVID-19 detection and preliminary diagnosis by mining datasets provided by health officials from different countries suffer from misinterpretations. However, authorities should observe privacy, security to prevent uncertain future use. In some cases, while collecting information from individuals, privacy and security are not adequately maintained. In the future, these pools of information can backfire by containing people based on data, not allowing human beings to enjoy their freedom, getting hacked by criminals for carrying out malicious activities.

Conclusion:

The DHSS is critical for taking preventive measures to evade public health misconceptions. Precise use of technology has and will guide humans to control public health-related issues. Nonetheless, misuse or erroneous employment of public health data can be fatal and cannot be robust overnight. Only proper use of DHSS can ensure public trust and confidence; otherwise, divergent outcomes will be highly expected. Contact tracing is found to be an essential public health approach to fight the spread of COVID-19 pandemic. However, in the developing countries, caution is warranted to generalize the usability of contract tracing and health surveillance apps [10]. Further detailed studies should be done to assess, predict the impact of DHSSs during COVID-19 period, and future health disasters. Protecting individual information in this era of big data, AI, and IoT would be an immense challenge. Appropriate guidelines should be developed and followed for smooth conduction of DHSS now, and in future.

Funding: None

Competing interests: None declared

Ethical approval: Not required

Conflict of interest: None